\documentclass{elsart}

\usepackage{graphicx}

\usepackage{amssymb}

\begin{document}

\begin{frontmatter}



\title{Nuclear transparencies in relativistic A$(e,e'p)$ models}


\author[label1]{P. Lava},
\author[label2]{M.C. Mart\'{\i}nez},
\author[label1]{J. Ryckebusch}\ead{Jan.Ryckebusch@ugent.be},
\author[label2]{J.A. Caballero},
\author[label3]{J.M. Ud\'{\i}as}

\address[label1]{Department of Subatomic and Radiation Physics, Ghent University, Proeftuinstraat 86, B-9000 Gent, Belgium}
\address[label2]{Departamento de F\'\i sica At\'omica, Molecular y Nuclear, Universidad de Sevilla, Apdo. 1065, E-41080 Sevilla, Spain}
\address[label3]{Departamento de F\'\i sica At\'omica, Molecular y Nuclear, Facultad de Ciencias F\'\i sicas, 
Universidad Complutense de Madrid, E-28040 Madrid, Spain
}

\begin{abstract}
Relativistic and unfactorized calculations for the nuclear
transparency extracted from exclusive $A(e,e'p)$ reactions for $0.3
\leq Q^2 \leq 10$~(GeV/c)$^2$ are presented for the target nuclei C, Si,
Fe and Pb.  For $Q^2 \geq 0.6$~(GeV/c)$^2$, the transparency results are
computed within the framework of the recently developed relativistic
multiple-scattering Glauber approximation (RMSGA).  The target-mass
and $Q^2$ dependence of the RMSGA predictions are compared with
relativistic distorted-wave impulse approximation (RDWIA)
calculations.  Despite the very different model assumptions 
underlying the treatment of the final-state interactions in the RMSGA
and RDWIA frameworks, they predict comparable nuclear transparencies
for kinematic regimes where both models are applicable.
\end{abstract}

\end{frontmatter}


The transparency of the nuclear medium to the propagation of protons
is an issue of fundamental importance. The nuclear transparency
provides a measure of the probability that a proton of a certain
energy escapes from the nucleus without enduring inelastic final-state
interactions (FSI). The nuclear transparency is a useful quantity for
studying nuclear medium effects, and in particular, it is very well
suited for investigations of the so-called color transparency
(CT) phenomenon, which predicts a significant enhancement of the
transmission of protons through nuclei
\cite{mueller82,brodsky82} once QCD mechanisms start playing a role. 

During the last decade, several investigations of the nuclear
transparency have been carried out using the $A(e,e'p)$ reaction in
the quasi-elastic (QE) regime. In this kinematic regime, the impulse
approximation (IA), where a quasifree single-nucleon knockout reaction
mechanism is assumed, has been proved to provide a good description of
the reaction dynamics. Thanks to the electromagnetic character of the
initial-state interactions in an $A(e,e'p)$ process, the entire nuclear
volume can be probed.

Nuclear transparency measurements with the $A(e,e'p)$ reaction are
available for a range of target nuclei. The first experiments were
performed at Bates for $Q^2 \approx 0.3$ (GeV/c)$^2$~\cite{Garino92},
and at SLAC for $1 \leq Q^2 \leq 7$
(GeV/c)$^2$~\cite{oneill95,makins94}.  Recently, measurements at the
Thomas Jefferson National Accelerator Facility (TJNAF) provided
precise data for the target nuclei $^2$D, $^{12}$C and $^{56}$Fe and
$Q^2=$ 3.3, 6.1 and 8.1 (GeV/c)$^2$ \cite{Garrow2002}. The same
facility provided an alternate set of data for the target nuclei $^{12}$C,
$^{56}$Fe and $^{197}$Au and 0.64$\leq Q^2 \leq$3.25 (GeV/c)$^2$
\cite{Abbott98,Dutta2003}.

The prediction of the nuclear transparency to protons poses a serious
challenge for models dealing with the $A(e,e'p)$ reaction due to the
wide range of proton energies which are probed in the present-day
experiments. As a matter of fact, at present there is no uniform and
realistic framework in which the proton-nucleus FSI effects can be
computed for proton kinetic energies ranging from 0.3 to several
GeV. For kinetic energies up to around 1 GeV, most theoretical
$A(e,e'p)$ investigations are performed within the context of the
Distorted-Wave Impulse Approximation (DWIA), where the effect of the
scatterings on the emerging nucleon is estimated with the aid of
proton-nucleus optical potentials. The parameterizations of these
potentials are usually not available for proton kinetic energies $T_p$
beyond 1 GeV. Beyond this energy, the Glauber model, which is a
multiple-scattering extension of the eikonal approximation, offers a
valid and economical alternative for describing FSI. In a Glauber
framework, the effects of FSI on the $A(e,e'p)$ observables are
computed directly from the elementary proton-nucleon scattering data
through the introduction of a profile function. The Glauber method
postulates linear trajectories and frozen spectator nucleons, and the
lower limit of this treatment to $A(e,e'p)$ has not yet been
established.

Numerous predictions for the nuclear transparencies within the context
of non-relativistic Glauber theory have been reported in literature
\cite{Frankel1994,nikolaev94,nikolaev95npa,Golubeva1998,frankfurt94a,frankfurt95,frankfurt01,pandharipande92,kohama93}. These
results are typically obtained in a
non-relativistic and factorized model for dealing with the
$e+A\rightarrow e' + (A-1)+ p$ reaction dynamics. In this context,
non-relativistic refers to the fact that the calculations use
bound-state wave functions or nuclear densities from solutions to a
Schr\"odinger equation and non-relativistic expressions for the
electromagnetic photon-nucleus interaction Lagrangian.  In the context
of modeling $A(e,e'p)$ processes, factorization refers to the
approximation of decoupling the electron-proton from the nuclear
dynamics part in the calculations.

In this paper we focus on relativistic and unfactorized descriptions
of nuclear transparencies extracted from quasi-elastic $A(e,e'p)$
processes.  In the past, relativistic distorted-wave impulse approximation (RDWIA) $A(e,e'p)$ calculations for the nuclear
transparency have been presented by Kelly \cite{Kelly1996} and Meucci
\cite{Meucci02}. Kelly used an effective current operator containing
the Dirac potentials, two-component bound states and distorted waves
obtained as solutions to relativized Schr\"odinger equations. Meucci used
bound-state wave functions from a relativistic mean-field approach,
while the effective Pauli reduction was adopted to construct the
ejectile's wave function.

For $T_p \geq 1$~GeV many partial waves need to be computed, thus a
description of the FSI mechanisms in terms of phenomenological optical
potentials may not be the most economical way at higher
energies. In addition, the global optical potentials we use in this work
were fitted to data with a limited range in $T_p$ below $1$~GeV, hence
the RDWIA studies of the nuclear transparency do not cover the
kinematic ranges beyond $T_p \mbox{=}1$~GeV.

Recently, a relativized and unfactorized version of the Glauber model
has been proposed \cite{debr02npa,debr02plb,ryckebusch03}. In this
paper, transparencies obtained within this so-called relativistic
multiple-scattering Glauber approximation (RMSGA) framework will be
compared with those obtained in the RDWIA framework as it has been
implemented by the Madrid-Sevilla group
\cite{Udias93,Udias95,Udias99,Udias01}. The comparison is made in a
consistent way.  For the transparency results which will be presented
below, this implies that the two frameworks only differ in the way
they treat the final-state interactions. All the remaining ingredients
are kept identical.

In what follows we will first sketch the basic ingredients entering
the RDWIA and the RMSGA frameworks, thereby indicating the similarities
and differences between the two.  Then, the energy dependence,
expressed in terms of the four-momentum transfer $Q^2= q^2 - \omega
^2$, and target-mass dependence of the nuclear transparencies obtained
in the two approaches will be compared and confronted with the data.

Adopting the IA, the central quantity to be computed in a relativistic
approach to $A(e,e'p)$ is the current matrix element
\begin{equation}
\langle J^{\mu}
\rangle = \int
d\vec{r} \; \overline{\phi}_F(\vec{r})\hat{J}^{\mu}(\vec{r})e^{i\vec{q}.\vec{
r}}\phi_B(\vec{r}) \; ,
\label{relcurrent}
\end{equation}
where $\phi_B$ and $\phi_F$ are relativistic wave functions describing
the initial bound and final outgoing nucleon respectively.  Further,
$\hat{J}^{\mu}$ is the relativistic one-body current operator.  The
bound-state wave function $\phi_B$ is a four-spinor with well-defined parity
and angular momentum quantum numbers ($\kappa_b$, $\mu_b$), obtained
within the framework of the relativistic independent-particle shell
model. It may be formally written as
\begin{equation}
\phi_B(\vec{r})=\left(\begin{array}{@{\hspace{0pt}}c@{\hspace{0pt}}}
                g_{\kappa_b}(r)\Phi_{\kappa_b}^{\mu_b}(\Omega_r) \\
                if_{\kappa_b}(r)\Phi_{-\kappa_b}^{\mu_b}(\Omega_r) \\
             \end{array}\right) \, ,
\label{bwf}
\end{equation}
with $\Phi_{\kappa_b}^{\mu_b}(\Omega_r)$ the usual spin spherical
harmonics.  The quantum numbers of the bound-state $\phi _B$ reflect
the nature of the single-hole state in which the residual nucleus is
left.

Within the RDWIA framework, $\phi_F$ is a scattering solution to a
Dirac-like equation, which includes scalar ($S$) and vector ($V$)
global optical potentials obtained by fitting elastic $proton-nucleus$ 
scattering data.  The scattering wave function, expressed in terms of
a partial-wave expansion in configuration space, reads
\begin{equation}
\phi_F({\vec{r}})=4\pi\sqrt{\frac{E_F+M}{2E_F}}\sum_{\kappa \mu
m}e^{-i \delta^\ast_\kappa} i^\ell\langle\ell m \frac{1}{2} s_F|j \mu
\rangle Y_\ell^{m\ast}(\Omega_{p_F})\Psi_\kappa^{\mu}({\vec{r}}) \, ,
\label{dwf}
\end{equation}
where $\Psi_\kappa^{\mu}({\vec{r}})$ are four-spinors of the same form
as in Eq.~(\ref{bwf}), except for the fact that they have complex
phase-shifts and radial functions. The outgoing nucleon momentum,
energy and spin are denoted as $\vec{p}_F$, $E_F$ and $s_F$ respectively.

In the RMSGA framework, the scattering wave function takes on the
following form 
\begin{equation}
\label{eq:transition}
\phi_F({\vec{r}}) \equiv  \phi _{p_F \; s_F}(\vec{r}) \mathcal{G} (\vec{b},z), 
\end{equation}
 where $\phi_{p_F \; s_F}$ is a relativistic plane wave.  The entire effect
of the FSI mechanisms is contained in the Dirac-Glauber phase
$\mathcal{G}(\vec{b},z)$, which is an A-body operator and reads
\begin{equation}
\label{eq:glauberphase}
\mathcal{G}(\vec{b},z)= \prod_{\alpha \neq B} \biggl[ 1-\int
d\vec{r}~'|\phi_{\alpha}(\vec{r}~')|^2 \theta(z'-z)\Gamma(\vec{b}
-\vec{b}') \biggr] ,
\end{equation} 
where the profile function for $pN$ scattering is defined as
\begin{equation}
\label{eq:profile}
\Gamma(\vec{b})=
\frac {\sigma_{pN}^{tot} (1-i\epsilon_{pN})} 
{4\pi\beta^2_{pN}} \exp(\frac{-b^2}{2\beta^2_{pN}})\;.
\end{equation}
The parameters $\sigma_{pN}^{tot}$, $\beta_{pN}$ and $\epsilon_{pN}$
depend on the proton kinetic energy and are obtained through
interpolation of the data base available from the Particle Data Group
\cite{pdg}.  We wish to stress that Glauber-based models come in very
different flavours and that the RMSGA formulation used here borrows a
lot of ingredients from the RDWIA approach to $A(e,e'p)$ processes,
except for the way of computing the effect of FSI mechanisms on the
hit proton, which is through-and-through different.

In order to make the comparisons between the RDWIA and RMSGA
transparency predictions as meaningful as possible, all the
ingredients in the $A(e,e'p)$ calculations not related to FSI, as
those concerning the implementation of relativistic dynamics and
nuclear recoil effects, are kept identical. In particular, both
pictures use the relativistic bound-state wave functions from a
Hartree calculation with the W1 parameterization for the different
field strengths \cite{furnstahl97}. Further, all the results presented
in this work are obtained within the Coulomb gauge using the so-called
$CC2$ current operator \cite{Forest1983}. For the description of
nuclear transparencies, the effect of Coulomb distortions has been
recognized as negligible \cite{Kelly1996}.  Therefore, no attempt has
been made to correct for the Coulomb-distortion effect.

We wish to stress that the RDWIA, as implemented by the Madrid-Sevilla group,
and RMSGA codes adopt very different numerical techniques to compute
the scattering wave functions and the corresponding matrix elements of
Eq.~(\ref{relcurrent}).  The Madrid RDWIA code employs a partial-wave
expansion to solve the Dirac equation for the ejectile.  The
cylindrical symmetry of the Glauber phase of
Eq.~(\ref{eq:glauberphase}) prohibits any meaningful use of this
technique in the RMSGA calculations.  Instead, the multi-dimensional
integrals are computed numerically.  In the limit of vanishing FSI
mechanisms, however, the two codes should predict identical results.
In the Glauber approach this limit is reached by putting the Glauber
phase of Eq.~(\ref{eq:glauberphase}) equal to unity.  In the RDWIA
picture, the effect of FSI can be made vanishing by
nullifying the optical potentials when solving the differential
equations for the radial wave functions determining the outgoing
nucleon wave functions. In this so-called relativistic plane-wave impulse approximation (RPWIA) limit, the Madrid code
is practically exact, as has been checked by comparing the
partial-wave expansion results with the analytical ones~\cite{Cab98}. In this plane-wave limit and assuming identical input options, the
RMSGA-Gent and RDWIA-Madrid codes produce differential cross sections
with an agreement to better than 5\%. This discrepancy can be
partially attributed to the numerical evaluation of the multi-dimensional integrals in RMSGA. This comparison gives us confidence about
the consistency of the calculations.

The nuclear transparency provides a measure of the likelihood that a
struck nucleon with kinetic energy $T_p$ escapes from the nucleus. The
nuclear transparency can be extracted from the measured $A(e,e'p)$
differential cross sections $ d^5 \sigma^{exp} (e,e'p) $ on the basis
of the following ratio
\begin{equation}
\label{eq:Texp}
T_{exp}(Q^2) = \frac{\int_{\Delta^3p_m}d\vec{p}_m\int_{\Delta E_m}dE_m
  \;S_{exp}(\vec{p}_m,E_m,\vec{p}_F)}{c_A
  \int_{\Delta^3p_m}d\vec{p}_m\int_{\Delta E_m}dE_m \;
  S_{PWIA}(\vec{p}_m,E_m)} \; .
\label{eq:texp}
\end{equation}
Here, $S_{exp}$ is the experimentally determined reduced cross
  section 
\begin{equation}
  S_{exp}(\vec{p}_m,E_m,\vec{p}_F) = \frac { \frac {d^5
  \sigma^{exp}} {d \Omega _p d \epsilon ' d \Omega _{\epsilon '}}
  (e,e'p) } {K \sigma _{ep} } , 
\end{equation}
 where $K$ is a kinematical factor and $\sigma _{ep}$ is the off-shell
 electron-proton cross section, which is usually evaluated with the
 $CC1$ prescription of de Forest \cite{Forest1983}. The quantities
 $\Delta^3p_m$ and $\Delta E_m$ specify the phase-space volume in the
 missing momentum and energy and are commonly defined by the cuts
 $|p_m| \leq 300$ MeV/c and $E_m \leq 80$ MeV. These kinematic cuts,
 in combination with the requirement that the Bjorken variable
 $x=\frac{Q^2}{2M_p \omega} \approx 1$, guarantee that the
 electro-induced proton-emission process is predominantly quasi-elastic.
 For example, the effects of two-body meson-exchange and isobar
 currents, which are neglected within the IA, have been shown to be
 at the percent level for quasi-elastic kinematics \cite{ryckebusch01,Ama03}.

In the above equation, $S_{PWIA}$ is the reduced cross section within
the plane-wave impulse approximation (PWIA) in the non-relativistic
limit. The factor $c_A$ in the denominator of Eq.~(\ref{eq:Texp}) has
been introduced to correct in a phenomenological way for short-range
mechanisms and is assumed to be moderately target-mass dependent.  It
accounts for the fact that short-range correlations move a fraction of
the single-particle strength to higher missing energies and momenta
and, hence, beyond the ranges covered in the integrations
$\int d\vec{p}_m \int dE_m$ of Eq.~(\ref{eq:texp}).  The values for $c_A$
which are adopted to extract the transparency from the $A(e,e'p)$
measurements are 0.9 ($^{12}$C), 0.88 ($^{28}$Si), 0.82 ($^{56}$Fe)
and 0.77 ($^{208}$Pb).

Theoretically, the nuclear transparencies are extracted from the
computed relativistic $A(e,e'p)$ angular cross sections for the
individual single-particle states, according to
\begin{equation}
\label{eq:Ttheo}
T_{theo}(Q^2)= \frac{\sum_{\alpha} \int_{\Delta^3p_m} d\vec{p}_m
  S^{\alpha} (\vec{p}_m,E_m,\vec{p}_F)} {c_A
  \sum_{\alpha}\int_{\Delta^3p_m} d\vec{p}_m S^{\alpha}_{PWIA}
  (\vec{p}_m,E_m)} \; .
\end{equation}
This expression reflects the one used to determine
$T_{exp}$.  Indeed, in our approach, we obtain the ``theoretical''
transparencies by adopting identical expressions and cuts as in the
experiments. Essentially, we replace the measured $A(e,e'p)$ angular
cross sections by the computed ones. In addition, the integration over
the missing energy $\int _{\Delta E_m} dE_m $ has been substituted by a
sum over all occupied shells ($\sum_{\alpha}$) in the ground
state of the target nucleus.  Indeed, the relativistic Hartree
approximation does predict bound-state eigenfunctions with a fixed
energy-eigenvalue and zero width. When determining the denominator in
Eq.~(\ref{eq:Ttheo}), in our calculations the PWIA limit is accomplished by nullifying all sources of FSI mechanisms and neglecting those contributions introduced by the presence of negative-energy components in the relativistic bound nucleon wave function \cite{Cab98}.

Transparencies have been calculated for the nuclei $^{12}$C,
$^{28}$Si, $^{56}$Fe and $^{208}$Pb. All numerical calculations are
performed in planar and constant $(q,\omega)$ kinematics. The adopted
values for $q$ and $\omega$ are the central values of the kinematics
in the $A(e,e'p)$ transparency experiments reported in Refs.~
\cite{Garino92,oneill95,Abbott98,Dutta2003}. For each shell $\alpha$,
the kinetic energy of the outgoing nucleon is calculated by means of
the relationship $T_p=\omega + \epsilon _ {\alpha}$, where $
\epsilon _ {\alpha}$ is the energy eigenvalue of the corresponding
single-particle state.  Due to the internal motion of the confined
protons, the ejected protons emerge in a cone about the transferred
momentum.  The boundaries of the cone are restricted by the
requirement that the ``initial '' proton momentum $|p_m| \leq 300$ MeV/c.

In the RDWIA calculations, we have employed the global $S-V$
parameterizations of Cooper et al~\cite{Cooper93}, which provide the
best phenomenological optical potentials to date. As the highest
kinetic energy in these parametrizations is 1 GeV, RDWIA
transparencies are obtained up to four-momentum transfers of $Q^2
\approx$ 1.8 (GeV/c)$^2$.  Due to its use of the eikonal
approximation, the validity of RMSGA becomes questionable when
approaching low values of $Q^2$. For this reason the RMSGA model is
not used for calculating transparencies below $Q^2 \approx$ 0.6
(GeV/c)$^2$.  Hence, the kinematic range $0.6 \leq Q^2 \leq
1.8$~(GeV/c)$^2$ will be covered in both the RMSGA and the RDWIA
frameworks.

\begin{figure}[p]
\begin{center}
\includegraphics[width=0.75\textwidth]{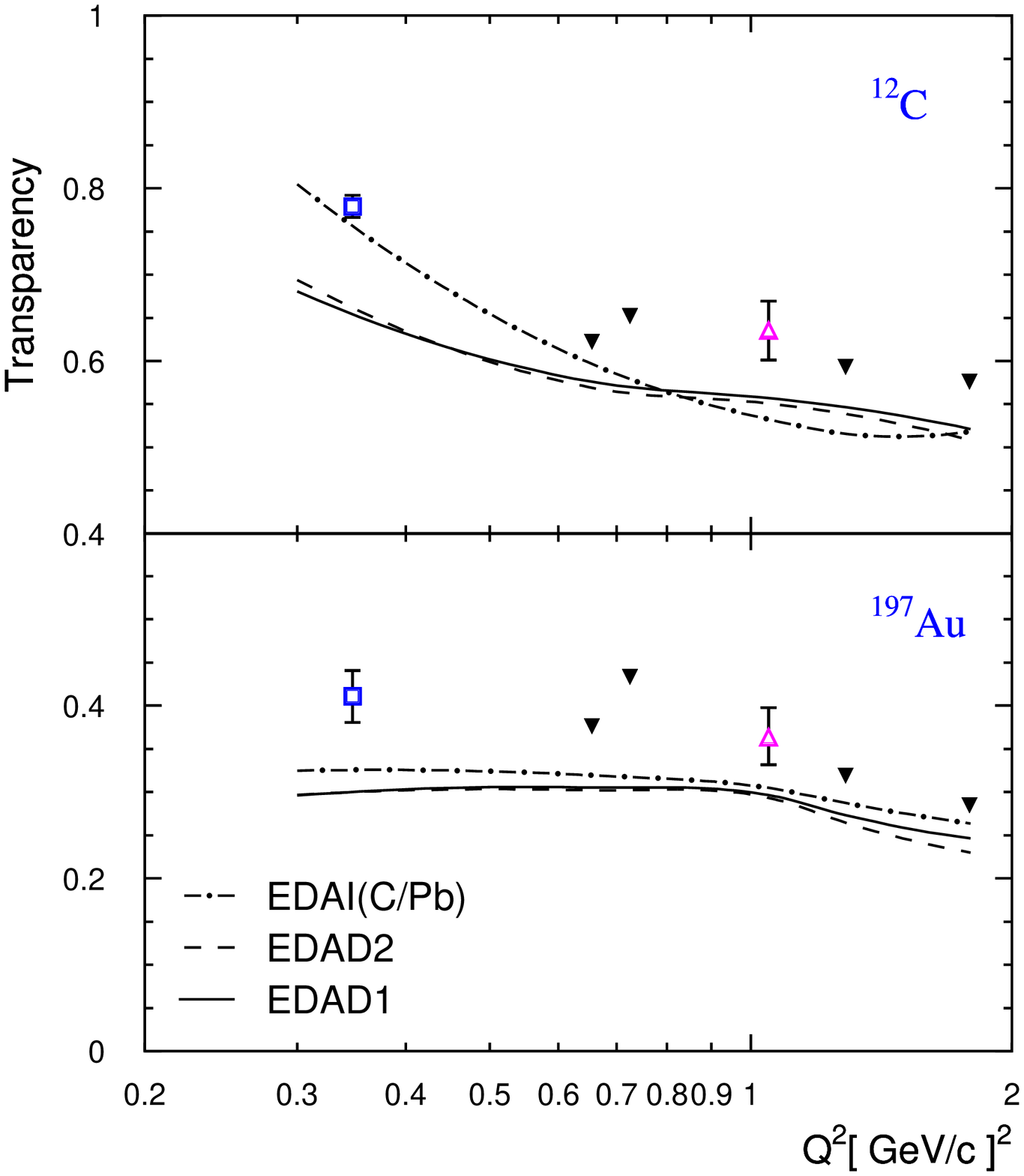}
\end{center}
\caption{The sensitivity of the computed nuclear transparencies in
  $^{12}$C and $^{197}$Au to the adopted choice for the
  parameterization of the relativistic optical potentials.  Results of
  RDWIA calculations with the EDAD1 (solid curve), EDAD2 (dashed
  curve) and EDAIC/EDAIPb (dot-dashed curve) are shown.  Data points are
  from Refs.~\cite{Garino92} (open squares), \cite{oneill95,makins94}
  (open triangles), and \cite{Abbott98,Dutta2003}(solid triangles)}
\label{fig.:transcompopt}
\end{figure}

First, we investigate the sensitivity of the computed transparencies
to the adopted parameterizations for the optical potentials.  In
Fig.~\ref{fig.:transcompopt} results for $^{12}$C and $^{208}$Pb are
displayed as a function of $Q^2$ for different optical-potential
parameterizations contained in Ref.~\cite{Cooper93}. For $^{12}$C,
both the predicted $Q^2$ dependence and the value of the transparency
depend on whether A-dependent (EDAD1/EDAD2) or A-independent (EDAIC)
fits for the potentials are selected. For $^{208}$Pb, the noted
differences between the different types of optical-potential sets are
less pronounced. Within the class of A-dependent parameterizations,
the versions EDAD1 and EDAD2 give rise to comparable nuclear
transparencies. In the remainder of the paper, the EDAD1 version will
be used. There are various arguments to motivate this choice. First,
the A-independent parameterization is only available for a very
limited number of nuclei, and extrapolation to other nuclei has been
discouraged \cite{Cooper93}.  Second, all energy-dependent A-dependent
parameterizations in Ref.~\cite{Cooper93} produce similar transparency
predictions. Finally, the relativistic transparency calculations by
Kelly~\cite{Kelly1996} and Meucci~\cite{Meucci02} employed the EDAD1
parametrization.  Adopting the same choice makes easier the comparison
between these predictions and ours.

\begin{figure}[p]
\begin{center}
\includegraphics[width=0.75\textwidth]{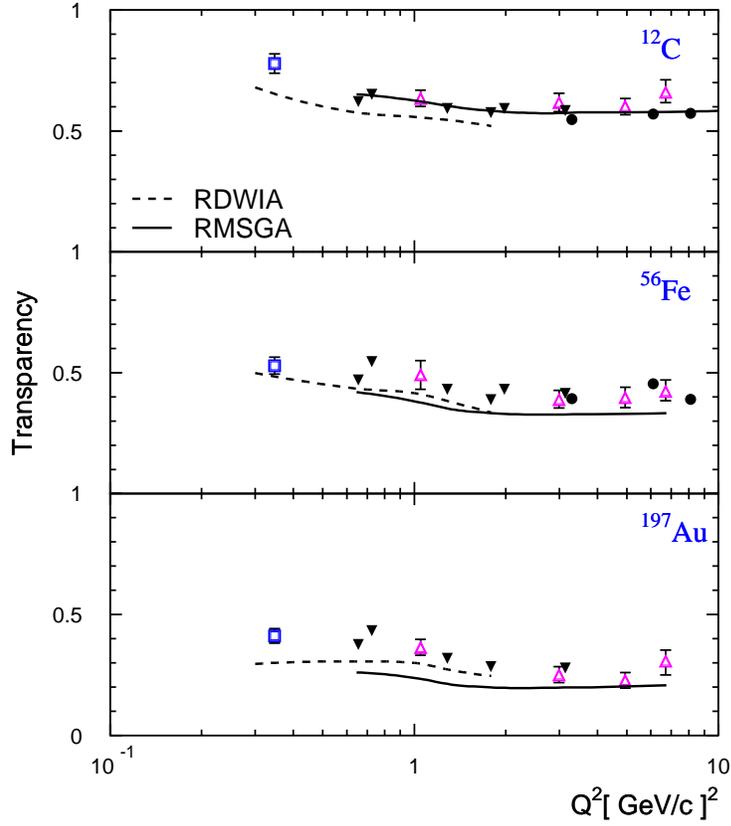}
\end{center}
\caption{Nuclear transparencies versus $Q^2$ for $A(e,e'p)$ reactions in
  quasi-elastic kinematics. The RMSGA (solid lines) are
  compared to the RDWIA (dashed lines) results. Data  are from
  Refs.~\cite{Garino92} (open squares), \cite{oneill95,makins94} (open
  triangles), \cite{Garrow2002}(solid circles) and
  \cite{Abbott98,Dutta2003}(solid triangles).}
\label{fig.:duttatrans}
\end{figure}

In Fig.~\ref{fig.:duttatrans}, the transparencies predicted by the
RMSGA and RDWIA models are displayed as a function of $Q^2$ and
compared to the world data. The $^{197}$Au data are compared to
$^{208}$Pb calculations. The RDWIA approach systematically
underestimates the data by roughly $5-10\%$.  The presented RDWIA
transparency results for $^{56}$Fe and $^{208}$Pb are in better
agreement with the data than those reported in
\cite{Abbott98}. The RDWIA transparencies obtained in
Ref.~\cite{Meucci02}, on the other hand, are rather comparable to
ours for low $Q^2$, the differences increasing for higher values.

The RMSGA framework reproduces the $^{12}$C data. With increasing
target mass, the agreement worsens, mostly for the lower values of
$Q^2$. The quality of agreement achieved for $^{12}$C and the
systematic underprediction of the transparencies for heavier nuclei,
was also a feature of the non-relativistic and factorized Glauber
calculations of Pandharipande and Pieper~
\cite{Garrow2002,pandharipande92}. A global feature of the RDWIA and
RMSGA calculations presented here, is that they tend to underestimate
the measured transparencies. Note that this is at variance with the
RDWIA calculations of \cite{Meucci02}.

As can be inferred from Fig.~\ref{fig.:duttatrans}, the RMSGA
framework predicts less absorption than RDWIA for a light nucleus like
$^{12}$C.  With increasing target mass the opposite holds true and
when approaching the heaviest target nuclei considered here, the
Glauber framework predicts 5 to 10 percent more absorption. The
measured $Q^2$ dependence is reasonably well reproduced by both
relativistic calculations. For low $Q^2$ the models reproduce the
trend of decreasing transparencies. For $Q^2 \geq 2$ (GeV/c)$^2$, the
RMSGA transparencies are close to constant, in line with the measured
ones and those predicted in typical non-relativistic Glauber
models. In fact, the modest energy variation of the transparency in
the RMSGA model is a reflection of the fact that the total and elastic
proton-nucleon cross sections remain fairly constant once $T_p \geq
1.7$~GeV.

In Ref.~\cite{Kelly1996} large discrepancies were observed between the
DWIA $A(e,e'p)$ transparencies and the ones from Glauber calculations
of Nikolaev \cite{nikolaev94,nikolaev95npa}.  In contradistinction,
Fig.~\ref{fig.:duttatrans} indicates reasonably good agreement between
our RDWIA and RMSGA model predictions for the medium-heavy nucleus
$^{56}$Fe and modest variations in opposite directions when moving to
a lighter or heavier nucleus. In Ref.~\cite{Kelly1996} the noted
differences between the transparencies obtained from DWIA and those
from the particular Glauber approach of
Refs.~\cite{nikolaev94,nikolaev95npa}, are attributed to the fact that
the latter adopts a closure property in deriving the expression for
the attenuation factor.  We wish to stress that this approximation is NOT used in the RMSGA formulation of Glauber theory. In
computing the effect of FSI mechanisms on the the $A(e,e'p)$ cross
sections, the sum extending over the occupied states $\alpha$ in
Eq.~(\ref{eq:Ttheo}) is carried out in a similar fashion in RMSGA and
RDWIA.

\begin{figure}[p]
\begin{center}
\includegraphics[width=0.75\textwidth]{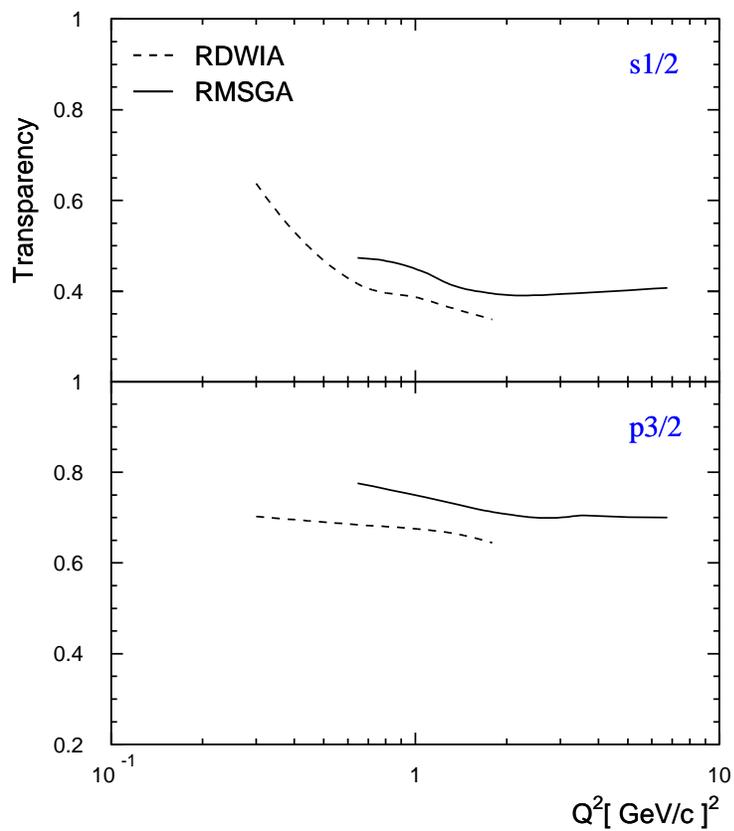}
\end{center}
\caption{The $Q^2$ dependence of the computed nuclear transparency for
the two single-particle orbits in $^{12}$C as obtained in the RDWIA
and RMSGA approach.}
\label{fig.:c12shelltransp}
\end{figure}

Investigating the attenuation for each individual shell in the target
nucleus allows one to study the radial dependence of the FSI
mechanisms.  In the $^{12}$C case, for example, the $1s_{1/2}$ has
spatial characteristics which are very different from the $1p_{3/2}$
orbit.  The attenuation for the individual states represents also a
more stringent test of the (non-)similarity of the optical-potential
and Glauber-based models for describing proton propagation through
nuclei. In Fig. \ref{fig.:c12shelltransp}, the RMSGA and RDWIA
predictions for the attenuation for the individual shells in $^{12}$C
are compared. These numbers are computed according to the definition
of Eq.~(\ref{eq:Ttheo}) without performing the sum over the states
$\alpha$.  Obviously, the optical-potential approach predicts more
absorption for both shells. As expected, both models predict a
stronger attenuation for proton emission from a level which has a
larger fraction of its density in the nuclear interior. Again, the
results of Fig.~\ref{fig.:c12shelltransp} illustrate that the
proton-nucleus (RDWIA) picture and the proton-nucleon picture (RMSGA)
are not dramatically different in their predictions.  These findings
provide us additional confidence that the ``low-energy'' and
``high-energy'' regime can be bridged in a relatively smooth
manner. Note further that the observed tendency of increasing $^{12}$C
transparencies at low $Q^2$, can almost be entirely attributed to the
$1s_{1/2}$ orbital.

\begin{figure}[p]
\begin{center}
\includegraphics[width=0.75\textwidth]{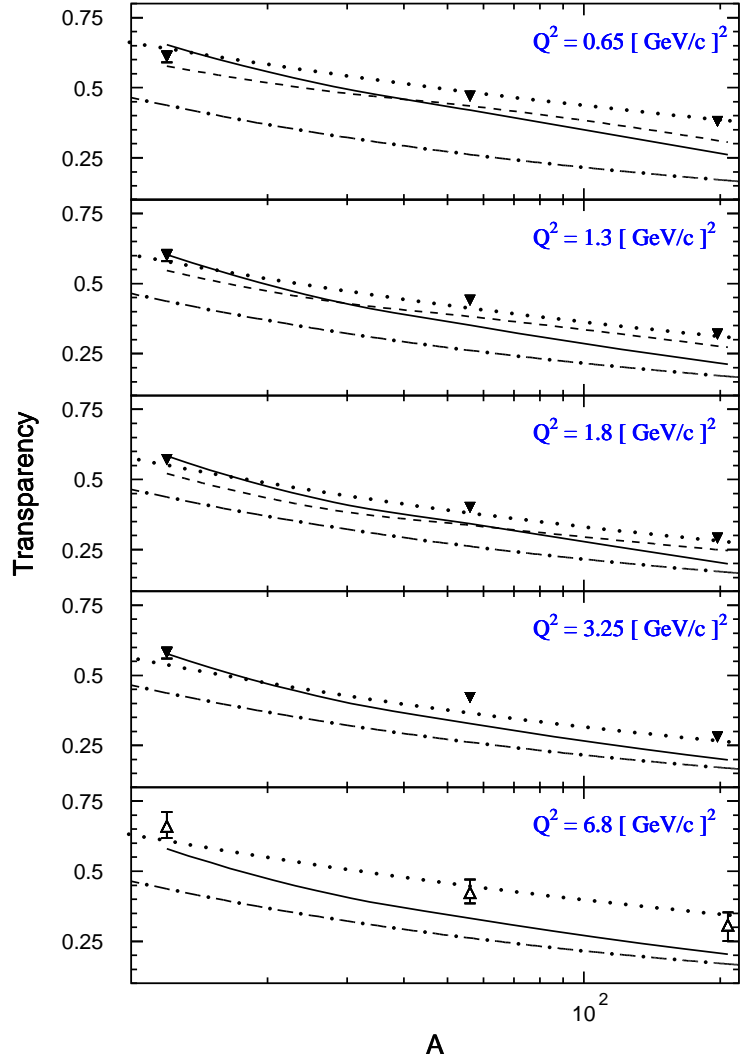}
\end{center}
\caption{The A-dependence of the nuclear transparency at five values
of the four-momentum transfer $Q^2$. The solid (dashed) curves are
RMSGA (RDWIA) calculations. The dotted curves represent the
$A^{-\alpha(Q^2)}$ parametrization, while the dot-dashed curve gives
$A^{-1/3}$. Data are from \cite{Abbott98,Dutta2003}(solid triangles)
and \cite{oneill95,makins94}(open triangles).}
\label{fig.:adep}
\end{figure}

The A-dependence of the nuclear transparencies at various values of
the four-momentum transfer is studied in Fig.~\ref{fig.:adep}.  The
RDWIA framework reproduces the measured A-dependence, while RMSGA
slightly overestimates it. Under the
assumption that the attenuation effect is proportional to the radius
of the target nucleus one would naively expect that the A-dependence
of the nuclear transparency can be parameterized as
\begin{equation}
T(Q^2) \mbox{=} c(Q^2)A^{-\alpha(Q^2)} \; ,
\end{equation} 
with $\alpha \mbox{=} 1/3$.  In the work of Ref.~\cite{Abbott98} it
was shown that the dependence of $T_{exp}(Q^2)$ on the mass number
could be nicely fitted with $c(Q^2)\equiv 1$ and $\alpha \equiv$ $0.17
\pm 0.04 (Q^2 \mbox{=} 0.65)$, $ 0.22 \pm 0.05 (Q^2 \mbox{=} 1.3)$, $ 0.24
\pm 0.04 (Q^2 \mbox{=} 1.8)$, $ 0.25 \pm 0.04 (Q^2 \mbox{=} 3.3)$, 
$0.20 \pm 0.02 (Q^2 \mbox{=} 6.8)$.  To guide the eyes these curves are
also displayed in Fig.~\ref{fig.:adep}.

In conclusion, we have presented for the first time a relativistic
calculation for the nuclear transparency for the process $e + A
\rightarrow e' + (A-1) + p$ covering the wide range of quasi-elastic
kinematics in which experiments have been performed.  An
optical-potential approach has been used up to the highest kinetic
energy ($T_p \approx 1$~GeV) for which potentials are readily
available.  Beyond that region we gathered our results within the
context of a relativized and unfactorized Glauber framework.  In a
medium-$Q^2$ range, both models have been applied and their
predictions compared.  Both frameworks accomodate relativistic effects
in the bound-state and scattering wave functions, as well as in the
electromagnetic current operator. Despite the very different
assumptions underlying the description of FSI effects in an
optical-potential and Glauber based approach to $A(e,e'p)$, their
predictions for the nuclear transparency and, in general, the effect
of attenuation for different single-particle levels, are comparable.

This work was partially supported by DGI (Spain) under Contracts Nos
BFM2002-03315, FPA2002-04181-C04-04, BFM2000-0600 and
BFM2003-04147-C02-01, by the Junta de Andaluc\'{\i}a and by
FWO-Flanders (Belgium) under Contract No
G.0020.03. M.C.M. acknowledges financial support from the Fundaci\'on
C\'amara (University of Sevilla).



\end{document}